\documentstyle[12pt,epsfig]{article}
\newcommand{\PL}[3]{{Phys. Lett.}        {#1} {(19#2)} {#3}}

\newcommand{\PR}[3]{{Phys. Rev.}        {#1} {(19#2)} {#3}}
\newcommand{\NP}[3]{{Nucl. Phys.}        {#1} {(19#2)} {#3}}
\def\ZPC{Z.\ Phys.\ C}

\newcommand{\czdot}{\! \cdot \!}
\newcommand{\beq}{\begin{equation}}
\newcommand{\eeq}{\end{equation}}
\newcommand{\beqa}{\begin{eqnarray}}
\newcommand{\eeqa}{\end{eqnarray}}
\newcommand{\beqan}{\begin{eqnarray*}}
\newcommand{\eeqan}{\end{eqnarray*}}
\newcommand{\ba}{\begin{array}}
\newcommand{\ea}{\end{array}}
\newcommand{\ben}{\begin{enumerate}}
\newcommand{\een}{\end{enumerate}}
\newcommand{\bfl}{\begin{flushleft}}
\newcommand{\efl}{\end{flushleft}}
\newcommand{\btab}{\begin{tabular}}
\newcommand{\etab}{\end{tabular}}
\newcommand{\bit}{\begin{itemize}}
\newcommand{\eit}{\end{itemize}}
\newcommand{\bdes}{\begin{description}}
\newcommand{\edes}{\end{description}}
\newcommand{\bdm}{\begin{displaymath}}
\newcommand{\edm}{\end{displaymath}}

\newcommand{\no}{\nonumber}

\newcommand{\ra}{\rightarrow}

\newcommand{\ve}{\varepsilon}

\newcommand{\dg}{\dagger}
\newcommand{\wt}{\widetilde}

\newcommand{\cL}{{\cal L}}
\newcommand{\M}{{\cal M}}

\newcommand{\cO}{{\cal O}}

\newcommand{\dfrac}{\displaystyle \frac}

\newcommand{\del}{\partial}

\begin{document}
\begin{titlepage}
\begin{flushright}
INFNNA-IV-96/47\\
UWThPh-1996-55\\
LNF-96/070 (P) \\
Dec. 1996\\
\end{flushright}
\begin{center}
 
\vspace*{1cm}
 
{\large \bf $\mbox{\boldmath $K \ra \pi \pi \pi \gamma$}$ in Chiral 
Perturbation Theory*} 
 
\vspace*{1cm}
{\bf{G. D'Ambrosio$^1$, G. Ecker$^2$, G. Isidori$^3$ and H. Neufeld$^2$}}
 
\vspace{.5cm}
${}^{1)}$ INFN, Sezione di Napoli \\
Dipartimento di Scienze Fisiche, Universit\`a di Napoli\\
I--80125 Napoli, Italy \\[5pt]
 
${}^{2)}$ Institut f\"ur Theoretische Physik, Universit\"at Wien\\
Boltzmanngasse 5, A--1090 Wien, Austria \\[5pt]
 
${}^{3)}$ INFN, Laboratori Nazionali di Frascati \\ 
P.O. Box 13, I--00044 Frascati, Italy

\vfill
{\bf Abstract} \\
\end{center}
\noindent
We present a complete analysis of $K \ra 3 \pi \gamma$ decays 
to $\cO (p^4)$ in the low--energy expansion of the Standard Model. 
We employ the notion of ``generalized bremsstrahlung'' to
take full advantage of experimental information on
the corresponding non--radiative $K \ra 3 \pi$ decays.

\vfill
\noindent * Work supported in part
by HCM, EEC--Contract No. CHRX--CT920026 (EURODA$\Phi$NE) and
by FWF (Austria), Project Nos. P09505--PHY, P10876-PHY
\end{titlepage}

\renewcommand{\thesection}{\arabic{section}}
\renewcommand{\thesubsection}{\arabic{section}.\arabic{subsection}}
\renewcommand{\theequation}{\arabic{section}.\arabic{equation}}
\setcounter{equation}{0}
\setcounter{subsection}{0}
 
\section{Introduction}
\label{sec:intro}

The present experimental status of $K\to 3 \pi \gamma$ decays
is rather meager. So far, only the two channels with a charged kaon
in the initial state have been detected experimentally with
very low statistics \cite{Bolotov,Barmin,Stamer}. None of the decay 
modes of a neutral kaon have been seen.

This unsatisfactory experimental situation will change soon, especially
after the completion of the $e^+ e^-$ collider DA$\Phi$NE in Frascati.
In this $\Phi$--factory one expects \cite{DAFNE} a total yield
of $7.5 \cdot 10^9 \; K_L K_S$ pairs and $1.1 \cdot 10^{10} \; K^+ K^-$
pairs per year. For up--to--date information on the future
prospects of kaon physics we refer to Ref.~\cite{wshop}.

With a sufficient number of events, what can one learn from a study
of those decays? The appropriate framework for such an investigation is
chiral perturbation theory \cite{CHPT} (CHPT). To
lowest order in an expansion in momenta and meson masses,
the radiative decays are completely determined \cite{DEIN96a} 
by the non--radiative amplitudes for $K\to 3 \pi$.
At next--to--leading order, a full--fledged CHPT calculation
of nonleptonic weak amplitudes of $\cO (p^4)$ is required (cf., e.g., 
Ref.~\cite{DEIN95}). Among other ingredients to be discussed in 
Sec.~\ref{sec:GB}, important components are the one--loop amplitudes with 
a single vertex from the lowest--order nonleptonic weak Lagrangian 
$\cL_2^{|\Delta S|=1}$ and tree--level amplitudes due to the corresponding 
Lagrangian $\cL_4^{|\Delta S|=1}$ of $\cO (p^4)$. 

There are three main issues we want to address:
\ben
\item[i.] Bremsstrahlung completely determines the lowest--order
amplitude, but it also contributes at next--to--leading order
(and at higher orders as well). Is there a unique procedure
to use all the available information on the non--radiative amplitudes
to $\cO (p^4)$, either from experiment or from theory?
The answer is positive as shown previously for a general radiative 
four--meson process \cite{DEIN96a}. Here, we put the concept
of ``generalized bremsstrahlung" \cite{DEIN96a} to a practical test.
\item[ii.] The nonleptonic weak Lagrangian of $\cO (p^4)$ contains a number
of low--energy constants \cite{KMW90,EKW93} that are little known at 
present. Can we expect to extract relevant information on those 
constants from $K\to 3 \pi\gamma$ data?
\item[iii.] More generally, can one make definite predictions for these 
radiative kaon decays within the Standard Model?
\een

The outline of the paper is as follows. In Sec.~\ref{sec:GB}, we
set up the kinematics and discuss the low--energy expansion of 
$K\to 3\pi$ and $K\to 3\pi \gamma$ amplitudes up to $\cO (p^4)$.
We discuss the concept of generalized bremsstrahlung 
that takes full advantage of the available experimental information 
on the non--radiative amplitude in the form of a
fourth--order polynomial in the momenta. In Sec.~\ref{sec:elamp},
we calculate the electric tree--level amplitude of $\cO (p^4)$ in terms
of the appropriate low--energy constants. We give a fairly complete
list of experimentally accessible radiative kaon decays that depend
on those weak constants of $\cO (p^4)$. The calculation of the electric
loop amplitude is deferred to an Appendix. To the same order in the chiral
expansion, the magnetic amplitude is a pure tree--level amplitude that
receives both direct (local) and reducible (nonlocal)
contributions. These are put together in Sec.~\ref{sec:magamp}.
Numerical results for rates and spectra of the four transitions
occurring at $\cO (p^4)$ are collected in Sec.~\ref{sec:num}. Some
conclusions are presented in Sec.~\ref{sec:conc}. All relevant formulas
for the one--loop amplitudes are contained in an Appendix,
recapitulating and applying the results of Ref.~\cite{DEIN96a}.

\setcounter{equation}{0}
\setcounter{subsection}{0}

\section{Low--energy expansion}
\label{sec:GB}
The kinematics of the decay 
$K(-p_4) \ra \pi_1(p_1) \pi_2(p_2) \pi_3(p_3) \gamma(k)$
is specified by five scalar variables which we choose as
\beq
s = (p_1 + p_2)^2~, \qquad \nu = p_4(p_1 -p_2)~, \qquad
t_i = k \czdot p_i \qquad (i = 1,\ldots,4) \label{kinema}
\eeq
with
$$
\sum_{i=1}^4 p_i + k = 0 ~, \qquad 
\sum_{i=1}^4 t_i = 0~.
$$
Any three of the $t_i$ together with $s$ and $\nu$ form a
set of independent variables.

The transition amplitude  
can be decomposed into an electric and a magnetic part:
\beq
\label{ampl}
A(K \ra 3 \pi \gamma) = e \ve^\mu(k)(E_\mu + 
\ve_{\mu\nu\rho\sigma} M^{\nu\rho\sigma})
\eeq
with
$$
k^\mu E_\mu = 0~, \qquad \ve_{\mu\nu\rho\sigma} k^\mu M^{\nu \rho\sigma}
= 0~.
$$


To lowest order in the chiral expansion, the amplitudes for both radiative
and non--radiative transitions are generated at tree level by the effective 
chiral Lagrangian of $\cO (p^2)$,
\beq \label{Leff2}
\cL_2 + \cL_2^{|\Delta S|=1}~.
\eeq
The strong part has the well--known form \cite{CHPT} 
\beq
\cL_2 = \frac{F^2}{4} \, \langle D_\mu U D^\mu U^\dg + 2B \M
(U + U^\dg)\rangle \label{L2}
\eeq
where $\langle \;\rangle$ denotes the trace in three--dimensional flavour
space.
$F$ is the pion decay constant in the chiral limit ($F
\simeq F_\pi = 92.4~\mbox{MeV}$), 
$\M$ is the quark mass matrix and $B$ is related to the quark condensate.
The unitary $3 \times 3$ matrix field $U$ incorporates the eight
pseudoscalar meson fields. In the exponential parametrization,
$$
U = \exp (i \sqrt{2} \Phi/ F)~,
$$
\beq
\Phi = \Phi^\dg = \left( \ba{ccc}
\dfrac{\pi_0}{\sqrt{2}} + \dfrac{\eta}{\sqrt{6}} & \pi^+ & K^+ \\[10pt]
\pi^- & - \dfrac{\pi_0}{\sqrt{2}} + \dfrac{\eta}{\sqrt{6}} & K^0 \\[10pt]
K^- & \bar K^0 & - \dfrac{2 \eta}{\sqrt{6}}
\ea \right)~, \label{conv}
\eeq
with $K_L = K^0_2 = (K^0 + \bar K^0)/\sqrt{2}$ and
$K_S = K^0_1 = i (K^0 - \bar K^0)/\sqrt{2}$ in the limit of CP conservation. 
For the processes under consideration, the covariant derivative 
$D_\mu U$ can be restricted to
$$
D_\mu U = \del_\mu U + i e A_\mu [Q,U]
$$
with the photon field $A_\mu$ and the quark charge matrix $Q$.

The weak $|\Delta S| =1$ Lagrangian in (\ref{Leff2}) can be written
in the form (our notation and conventions are those of Ref.~\cite{DEIN95})
\beq
\cL_2^{|\Delta S|=1} = G_8 F^4 \langle \lambda L_\mu L^\mu\rangle +
G_{27} F^4 \left( L_{\mu 23} L^\mu_{11} + \frac{2}{3} L_{\mu 21}
L^\mu_{13}\right) + {\rm h.c.}~, \label{L2w}
\eeq
$$
\lambda = \frac{1}{2} (\lambda_6 - i \lambda_7)~, \qquad
L_\mu = i U^\dg D_\mu U~.
$$
The coupling constants $G_8$, $G_{27}$ in (\ref{L2w}) measure
the strength of the octet and the 27--plet part, respectively, of the
strangeness changing weak interactions. From $K \ra \pi \pi$
decays one finds
\beq
|G_8| \simeq 9 \cdot 10^{-6} \mbox{ GeV}^{-2}~, \qquad
G_{27}/G_{8} \simeq 1/18~.
\eeq

At lowest order, the magnetic amplitude $M^{\nu \rho\sigma}$ 
in (\ref{ampl}) vanishes since
there is no $\ve$ tensor in the Lagrangian (\ref{Leff2}). The electric
amplitude, on the other hand, is completely determined by the corresponding
non--radiative amplitude $A(s,\nu)$ via Low's theorem \cite{Low58}:
\beqa
\label{Low}
E^\mu &=& A(s,\nu) \Sigma^\mu \no \\*
&& \mbox{} + 2 \frac{\partial A(s,\nu)}{\partial s} \Lambda^\mu_{12} +
\frac{\partial A(s,\nu)}{\partial \nu} (\Lambda^\mu_{14} -
\Lambda^\mu_{24}) \no \\*
&& \mbox{} + \cO(k)
\eeqa
with (the meson charges in units of $e$ are denoted $q_i$, with
$\sum_{i=1}^4 q_i = 0$)
\beqa
\label{defs}
\Sigma^\mu &=& \sum_{i=1}^4 \frac{q_i p_i^\mu}{t_i} \no \\
\Lambda^\mu_{ij} &=& \Lambda^\mu_{ji} = (q_i t_j - q_j t_i) D^\mu_{ij}
\no \\
D^\mu_{ij} &=& - D^\mu_{ji} = \frac{p_i^\mu}{t_i} - \frac{p_j^\mu}{t_j}~.
\eeqa
Since there are no terms of $\cO(k)$ at lowest order in the chiral expansion, 
the leading--order electric amplitude is completely determined by the 
explicit terms in (\ref{Low}) usually called ``internal bremsstrahlung".
 
At next--to--leading order, $\cO (p^4)$, the situation is much more
complicated. A nonleptonic weak amplitude of $\cO (p^4)$ receives
in general four types of contributions \cite{DEIN95}:
\ben
\item[i.] Tree--level amplitudes from the effective chiral
Lagrangian $\cL_4^{|\Delta S|=1}$ of $\cO(p^4)$ with the proper octet and
27--plet transformation properties.
\item[ii.] One--loop amplitudes from diagrams with a single vertex
from $\cL_2^{|\Delta S|=1}$ in the loop.
\item[iii.] Reducible tree--level amplitudes with a single vertex from
$\cL_2^{|\Delta S|=1}$ and with a single vertex either from the strong 
Lagrangian $\cL_4$ or from the anomalous Wess--Zumino--Witten
Lagrangian \cite{WZW71}.
\item[iv.] Reducible one--loop amplitudes, consisting
of a strong loop diagram connected to a vertex of $\cL_2^{|\Delta S|=1}$
by a single meson line. A typical diagram of this type
contains an external $K-\pi$ or $K-\eta$ transition, possibly
with an additional photon (generalized ``pole diagrams''). The calculation
of such diagrams is simplified by a rediagonalization of the kinetic
and mass terms of $\cL_2 + \cL_2^{|\Delta S|=1}$ (``weak rotation''
\cite{EPRbc}).
\een

For the decays $K\to 3 \pi\gamma$, all four mechanisms are relevant. Most
of them also appear in the non--radiative amplitudes. Via
Low's theorem (\ref{Low}), the non--radiative amplitude of $\cO (p^4)$
will contribute to the electric part of the radiative amplitude. 
Unlike at lowest order, this is however not the whole story at
$\cO (p^4)$. The question then is how to use in an optimal way the amplitude 
$A(s,\nu)$ of $\cO (p^4)$, either from theory or from experiment,
for calculating the radiative electric amplitude $E^\mu$
of the same order.

In a recent paper \cite{DEIN96a}, we have presented
the general theoretical framework for the treatment of radiative four--meson
amplitudes like $K \ra 3 \pi \gamma$.
The essential point is the concept of ``generalized bremsstrahlung'',
\beq
E^\mu = E^\mu_{\rm GB} + \cO(k)~, 
\eeq
where $E^\mu_{\rm
GB}$ is defined in terms of the non--radiative  amplitude $A(s, \nu)$ 
\cite{DEIN96a}:
\beqa
\label{EGB}
E^\mu_{\rm GB} &=& A(s,\nu) \Sigma^\mu + 
2 \frac{\partial A(s,\nu)}{\partial s} \Lambda^\mu_{12} + 
\frac{\partial A(s,\nu)}{\partial \nu}(\Lambda^\mu_{14} - \Lambda^\mu_{24})
\no \\
&& \mbox{} +
2 \frac{\partial^2 A(s,\nu)}{\partial s^2} (t_1 + t_2)\Lambda^\mu_{12} + 
\frac{1}{2}
\frac{\partial^2 A(s,\nu)}{\partial \nu^2} [(t_1 - t_2)
(\Lambda^\mu_{14} - \Lambda^\mu_{24}) - t_3 t_4 \Sigma^\mu] \no \\
&& \mbox{} + 2 \frac{\partial^2 A(s,\nu)}{\partial s \partial \nu}
[t_2 \Lambda^\mu_{14} - t_1 \Lambda^\mu_{24}]~.
\eeqa

Referring to Ref.~\cite{DEIN96a} for a more thorough exposition,
we concentrate here on the practical advantages of generalized
bremsstrahlung. Many of the terms in the above list of
four mechanisms appear in both the radiative and the non--radiative
amplitudes and are therefore automatically included in $E^\mu_{GB}$.
This is in particular true for most of the renormalization parts
that are trivially carried over from $A(s,\nu)$ to $E^\mu$, but also
for many of the so--called reducible contributions (items iii and iv
in the above list). For instance, all the weak low--energy constants
$N_i$ \cite{EKW93} contributing to both $K\to 3 \pi$ and 
$K\to 3 \pi\gamma$ are completely taken into account by $E^\mu_{GB}$.
Therefore, only the genuine radiative low--energy constants $N_{14}$,
\dots, $N_{17}$ will show up in $E^\mu - E^\mu_{GB}$. 

In the following we use the experimental  
$K \ra 3 \pi$ amplitudes to derive $E^\mu_{\rm GB}$. 
If we had limited ourselves to an analysis at the center of the 
Dalitz plot of  $K\rightarrow 3\pi$ data \cite{Dalitz,Handbook1,Sannino}
or just to linear slopes \cite{Ferrari},
there would have been no need to extend (\ref{Low}) to (\ref{EGB}). 
However, the quadratic slopes are observed and the $K \ra 3 \pi$ amplitudes
are written as polynomials of second order in $s$ and $\nu$
\cite{DD79,KMW91,DIPP94} to fit the experimental data. The second 
derivatives in (\ref{EGB}) are thus needed to take advantage of all 
the  experimental information available from  $K\rightarrow 3\pi$. 
The (electric) direct emission term $E^\mu - E^\mu_{\rm GB}$ is then a 
genuine radiative part of the amplitude not related to the non--radiative 
transition.

In the numerical analysis we have used the following parametrization of 
the $K \ra 3 \pi$ amplitudes
\cite{DIPP94,Maiani95}:
\beqa
\label{par}
A(K^+ \ra \pi^0 \pi^0 \pi^+) &=& a_c (1 + i\alpha_0 -
i\alpha'_0Y) - [b_c (1 + i\beta_0) - b_2 (1 + i\delta_0)]Y
\no \\
&& \mbox{} + c_c (Y^2 + X^2/3) - (d_c - d_2)(Y^2 - X^2/3)~,
\no \\
A(K^+ \ra \pi^+ \pi^+ \pi^-) &=& 2 a_c (1 + i\alpha_0 +
i\alpha'_0Y/2) + [b_c (1 + i\beta_0) + b_2 (1 + i\delta_0)]Y
\no \\
&& \mbox{} + 2 c_c (Y^2 + X^2/3) + (d_c + d_2)(Y^2 - X^2/3)~,
\no \\
A(K_L \ra \pi^+ \pi^- \pi^0) &=& a_n (1 + i\alpha_0 -
i\alpha'_0Y) - b_n (1 + i\beta_0)Y
\no \\
&& \mbox{} +  c_n (Y^2 + X^2/3) - d_n (Y^2 - X^2/3)~,
\no \\
A(K_S \ra \pi^+ \pi^- \pi^0) &=& - 2 i[ b_2 (1 + i\delta_0)
- 2 d_2 Y]X/3~,
\eeqa
with
\beq
X = 2 \nu / M_{\pi^+}^2~, \qquad Y = (s -s_0) / M_{\pi^+}^2~, \qquad
s_0 = \sum_{i=1}^4 M_i^2 /3~.
\eeq
The numerical values for $a_c$, $b_c$, etc. (in units of $10^{-8}$) are
given by \cite{KMW91,Maiani95}:
\beq
\ba{ll}
a_c = -95.39 \pm 0.40~, \qquad & a_n = 84.35 \pm 0.57~, \\
b_c =  24.47 \pm 0.34~, \qquad & b_n = -28.11 \pm 0.49~, \\
c_c =   0.68 \pm 0.17~, \qquad & c_n = - 0.05 \pm 0.22~, \\
d_c = - 1.63 \pm 0.34~, \qquad & d_n =   1.27 \pm 0.45~, \\
b_2 =  -3.91 \pm 0.40~, \qquad & d_2 =   0.21 \pm 0.51~.  
\ea \label{numval}
\eeq
For the phases associated with the absorptive parts in
(\ref{par}) we use the lowest--order CHPT 
predictions $\alpha_0 = 0.13$, $\alpha'_0 = -0.12$
and $\beta_0 = -\delta_0 = 0.047$ \cite{DIPP94}.

The decomposition (\ref{par}) is based on isospin symmetry. Moreover,
the numerical values in (\ref{numval}) have been obtained by a 
fit \cite{KMW91} where for simplicity the imaginary parts were set to 
zero. Present data on $K\to 3 \pi$ are too poor\footnote{~Note that we 
have not taken into account the very recent and accurate results of 
Serpukhov-167~\protect\cite{Serpukhov} in the $K^+\to\pi^0\pi^0\pi^+$
channel.} (especially in the $K_S$ channel) 
both to relax the assumption of isospin conservation and to be sensitive to 
the small imaginary parts. As a consequence, our numerical predictions for 
the generalized bremsstrahlung amplitudes in  $K\to 3\pi\gamma$
are affected by systematic errors and must be considered as 
preliminary. A new detailed analysis should be performed when 
complete and accurate $K\to 3\pi$ data will be available.

In the next two sections we discuss separately the electric and
the magnetic amplitudes for the various channels. To $\cO (p^4)$, there are
four non--vanishing transitions:
\beqa
K^+\to\pi^0\pi^0\pi^+\gamma & \qquad & K^+\to \pi^+\pi^+\pi^-\gamma \no \\
K_L\to \pi^+\pi^-\pi^0\gamma & \qquad & K_S\to \pi^+\pi^-\pi^0\gamma \no~.
\eeqa
We make the following simplifications for the calculation. The 27--plet
part of the nonleptonic weak Lagrangian is 
not included in the calculation of direct emission amplitudes, i.e.
in $(E^\mu - E^\mu_{\rm GB})$ and $M^{\nu\rho\sigma}$. This is an excellent
approximation in view of the $\Delta I=1/2$ rule. Moreover, in the
loop diagrams we have only kept the dominant two--pion intermediate
states. Since the loop amplitudes will turn out to be rather small anyway,
this restriction is justified a posteriori. Finally,
CP conservation will be assumed throughout the analysis.

\setcounter{equation}{0}
\setcounter{subsection}{0}
 
\section{Electric amplitudes}
\label{sec:elamp}
To $\cO (p^4)$, the electric amplitude can 
be written as 
\beq
E^\mu = E^\mu_{\rm GB} + E^\mu_{\rm counter}+E^\mu_{\rm loop,subtracted} ~. 
 \label{Efin}
\eeq
Use of the generalized bremsstrahlung amplitude $E^\mu_{\rm GB}$
greatly simplifies the calculation of both the tree--level and the
loop part of (\ref{Efin}). For instance, all the reducible contributions
(items iii and iv in the list of Sec.~\ref{sec:GB}) to the
electric amplitude are automatically contained in $E^\mu_{\rm GB}$. 
This can be shown almost without any calculation by going
back to the definition (\ref{EGB}) of generalized bremsstrahlung. The only
exception that needs some (tree--level) calculations are amplitudes 
proportional to the strong low--energy
constant $L_9$ \cite{CHPT} with an external weak transition.
Although there are strong radiative four--meson amplitudes proportional
to $L_9$, the explicit calculation shows that they do not contribute to 
$K\to 3 \pi\gamma$ after a weak rotation.

Another consequence of using generalized bremsstrahlung in (\ref{Efin})
is a much simpler form of $E^\mu_{\rm counter}$. All the
low--energy constants appearing in both radiative and non--radiative
amplitudes are already contained in $E^\mu_{\rm GB}$. 
Therefore, only the genuine radiative terms in the octet Lagrangian of
$\cO (p^4)$ \cite{EKW93}
\beq
\cL_4^{|\Delta S| = 1} = G_8 F^2 \sum_i N_i W_i + {\rm h.c.}~,\label{L4w}
\eeq
with dimensionless coupling constants $N_i$ and octet operators $W_i$,
contribute to $E^\mu_{\rm counter}$. In particular, 
going through the Lagrangian (\ref{L4w}) one finds that only
the four low--energy constants $N_{14}$, \dots, $N_{17}$
can occur in $E^\mu_{\rm counter}$. 
The relevant parts of the Lagrangian
(\ref{L4w}) are listed below.

With $F_{\mu\nu} = \partial_{\mu} A_{\nu} - \partial_{\nu}
A_{\mu}$ the electromagnetic field strength tensor, the explicit coupling
for $K^+ \ra \pi^0 \pi^0 \pi^+ \gamma$ is given by 
\beq
-\frac{i e G_8}{F^2} (N_{14} - N_{15} - N_{16} -N_{17})
F_{\mu\nu} K^+ \pi^0  \partial^\mu \pi^0 \partial^\nu \pi^-~.
\eeq
The corresponding expression for $K^+ \ra \pi^+ \pi^+ \pi^- \gamma$ reads 
\beq
-\frac{4 i e G_8}{F^2} (N_{14} - N_{15} - N_{16} -N_{17})
F_{\mu\nu} K^+ \pi^-  \partial^\mu \pi^+ \partial^\nu \pi^-~.
\eeq
The decay $K_L \ra \pi^+ \pi^- \pi^0 \gamma$ receives a
contribution from 
\beq
-\frac{i e G_8}{F^2} (N_{14} - N_{15} - N_{16} -N_{17})
F_{\mu\nu} K_L (\partial^\mu \pi^0 \pi^-
\stackrel{\leftrightarrow}{\partial^\nu} \pi^+ - 2 \pi^0
\partial^\mu \pi^+ \partial^\nu \pi^-)~,
\eeq
and $K_S \ra \pi^+ \pi^- \pi^0 \gamma$ from 
\beq
-\frac{ e G_8}{F^2} [7 (N_{14} - N_{16}) + 5 ( N_{15} + N_{17})]
F_{\mu\nu} K_S \partial^\mu \pi^0 (\pi^- \partial^\nu \pi^+ +
\pi^+ \partial^\nu \pi^-)~.
\eeq
\begin{table}[t]
\caption{Kaon decay modes to which the coupling constants $N_i$ contribute.
For the $3 \pi$ final states, only the single photon channels are
listed. For the neutral modes, the letters $L$ or $S$ in brackets
distinguish between $K_L$ and $K_S$ in the limit of
CP conservation. $\gamma^*$ denotes a lepton pair in the final state.
If a decay mode appears more than once there are different Lorentz
structures in the amplitude. The combinations with $N_i^r$ are scale
dependent compensating the scale dependence of the corresponding loop
amplitude. The other combinations are scale independent.}
\label{tab:Ni}
$$
\begin{tabular}{|c|c|c|c|} \hline
$\pi$ & $2 \pi$ & $3 \pi$ & $N_i$ \\ \hline
$\pi^+ \gamma^*$ &$\pi^+ \pi^0 \gamma^*$ & & $N_{14}^r - N_{15}^r$\\
$\pi^0 \gamma^*~(S)$ & $\pi^0\pi^0\gamma^*~(L)$ & & $2 N_{14}^r + N_{15}^r$\\
$\pi^+ \gamma\gamma$ & $\pi^+\pi^0\gamma\gamma$ & & $N_{14} - N_{15}
-2 N_{18}$ \\
 & $\pi^+\pi^-\gamma\gamma~(S)$ & & " \\
 & $\pi^+\pi^0\gamma$ & $\pi^+\pi^+\pi^-\gamma$ & $N_{14}-N_{15}-N_{16}
-N_{17}$ \\
 & $\pi^+ \pi^- \gamma~(S)$ & $\pi^+\pi^0\pi^0\gamma$ & " \\
 & & $\pi^+\pi^-\pi^0\gamma~(L)$ & " \\
 & & $\pi^+\pi^-\pi^0\gamma~(S)$ & $7(N_{14}^r-N_{16}^r)+ 5(N_{15}^r
+ N_{17})$ \\
 & $\pi^+\pi^-\gamma^*~(L)$ & & $ N_{14}^r - N_{15}^r -3(N_{16}^r
-N_{17})$\\
 & $\pi^+\pi^-\gamma^*~(S)$ & & $ N_{14}^r - N_{15}^r -3(N_{16}^r
+N_{17})$\\
 & $\pi^+\pi^0\gamma^*$ & & $ N_{14}^r + 2 N_{15}^r -3(N_{16}^r
-N_{17})$\\
\hline
 & $\pi^+\pi^-\gamma~(L)$ & $\pi^+\pi^-\pi^0\gamma~(S)$ & $N_{29} + N_{31}$ \\
 & & $\pi^+\pi^+\pi^-\gamma$ & " \\
 & $\pi^+\pi^0\gamma$ & $\pi^+\pi^0\pi^0\gamma$ & $3 N_{29} - N_{30}$ \\
 & & $\pi^+\pi^-\pi^0\gamma~(S)$ & $5 N_{29}  -N_{30}+ 2 N_{31} $ \\
 & & $\pi^+\pi^-\pi^0\gamma~(L)$ & $6 N_{28} + 3 N_{29} - 5 N_{30}$
\\ \hline
\end{tabular}
$$ 
\end{table}

In order to facilitate the comparison with other radiative kaon decays,
we list in Table \ref{tab:Ni} the combinations of low--energy constants
$N_i$ governing the various experimentally accessible channels. This
Table is a slightly extended version of the one appearing in Ref.~\cite
{DEIN95}. As one can see from the Table, the specific combination of 
coupling constants $N_{14} - N_{15} - N_{16} -N_{17}$ occurs 
also in the amplitudes for $K^+ \ra \pi^+ \pi^0 \gamma$ and 
$K_S \ra \pi^+ \pi^-\gamma$. On the other hand, $7 (N_{14}^r - N_{16}^r) 
+ 5 ( N_{15}^r + N_{17})$ is a characteristic combination\footnote{We
remind the reader that $N_{17}$ is scale independent \cite{EKW93}.}
for $K_S \ra \pi^+ \pi^- \pi^0 \gamma$ only. 

Both combinations are not yet known phenomenologically. To get a feeling 
for the typical size of these couplings one may appeal to the
factorization model that predicts \cite{EKW93}
\beqa
[N_{14} - N_{15} - N_{16} - N_{17}]^{\rm FM} 
&=& - k_f \frac{F_\pi^2}{2 M_V^2} =
- 7 \cdot 10^{-3} k_f~, \label{counterterm1} \\
\left[7 (N_{14} - N_{16}) + 5 ( N_{15} + N_{17})\right]^{\rm FM} &=& 
41 k_f \frac{F_\pi^2}{2 M_V^2}~,\label{KScount}   
\eeqa
where $k_f$ is a fudge factor which naive factorization sets
equal to one. Note the potentially large counterterm amplitude in
$K_S \ra \pi^+ \pi^- \pi^0 \gamma$. Table \ref{tab:Ni} also indicates 
that the combination $N_{14} - N_{15} - N_{16} - N_{17}$ is scale independent
while $7 (N_{14}^r - N_{16}^r) + 5 ( N_{15}^r + N_{17})$ is not. 
Consequently, the loop amplitudes are all finite
for $K^+ \ra \pi^0 \pi^0 \pi^+ \gamma$,~$K^+ \ra \pi^+ \pi^+ \pi^- \gamma$
and $K_L \ra \pi^+ \pi^- \pi^0 \gamma$, but divergent
for $K_S \ra \pi^+ \pi^- \pi^0 \gamma$. This divergence is renormalized
by the counterterm combination $7 (N_{14} - N_{16}) 
+ 5 ( N_{15} + N_{17})$. In the limit where $G_{27}$ is set to zero,
the two--pion loop does not contribute to the $K_S$ decay . 
Since we have not included the other loop contributions
that are numerically negligible, the amplitude of $\cO (p^4)$
for the $K_S$ decay is superficially scale dependent. We shall come
back in Sec.~\ref{sec:num} to investigate numerically the effect of this
scale dependence.

\begin{figure}
\centerline{\epsfig{file=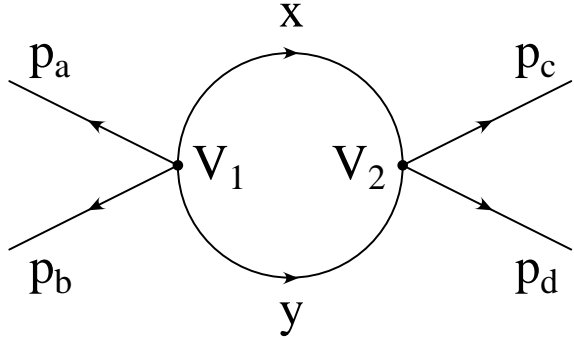,height=4cm}}
\caption{One--loop diagram for a four--meson transition.
For the radiative amplitude, the photon must be appended to every
charged meson line and to every vertex with at least two charged fields.
For the case of $K\to 3\pi(\gamma)$, $-p_a=-p_4$ is the kaon momentum,
the other three being the pion momenta. The weak (strong) vertex 
 $V_1$ ($V_2$) is defined in Eq.~(\protect\ref{vertex}), with the
appropriate coefficients for the various diagrams given in 
Table \protect\ref{tab:coeff}.}\label{fig:loop}
\end{figure}

Finally, the loop contributions to (\ref{Efin}) have to be
calculated. Once again, many contributions are already contained in
$E^\mu_{\rm GB}$. The only type of diagram that has to be calculated
explicitly is shown in Fig.~\ref{fig:loop} where a photon can be appended 
to all (charged) lines and vertices. In this diagram, $V_1$ is a weak
vertex from $\cL_2^{|\Delta S|=1}$ and $V_2$ is a strong vertex from
$\cL_2$. Of course, such diagrams without a photon contribute also to
the $K\to 3\pi$ amplitudes of $\cO (p^4)$. In accordance with the definition
of generalized bremsstrahlung in (\ref{EGB}), the appropriate part has
be subtracted from the radiative loop amplitude to obtain 
$E^\mu_{\rm loop,subtracted}$ in the complete amplitude (\ref{Efin}).

The calculation of the loop amplitudes is rather involved in the radiative
case. We have given in Ref.~\cite{DEIN96a} a compact expression for
the radiative loop amplitude with
general vertices $V_1,~V_2$ of $\cO (p^2)$. In an Appendix,
we reproduce the main steps for arriving at the final amplitude,
together with the relevant vertices for $K\to 3\pi\gamma$.

\setcounter{equation}{0}
\setcounter{subsection}{0}
 
\section{Magnetic amplitudes}
\label{sec:magamp}

The magnetic amplitude in (\ref{ampl}) receives contributions 
from direct and reducible diagrams \cite{ENP92,ENP94} corresponding
to type i and iii, respectively, in the classification of 
Sec.~\ref{sec:GB}.

The direct parts (type i) are
generated by the operators $W_{28}$, \dots, $W_{31}$ in (\ref{L4w}).
Their contribution to 
$K^+ \ra \pi^0 \pi^0 \pi^+ \gamma$ is given by 
\beq \label{ano1}
-\frac{ e G_8}{F^2} (3 N_{29} - N_{30})
\wt F_{\mu\nu} \partial^\mu K^+ \partial^\nu \pi^- \pi^0 \pi^0 ~,
\eeq
where $\wt F_{\mu\nu} = \ve_{\mu\nu\rho\sigma} F^{\rho\sigma}$ 
($\ve_{0123} = +1$). The
corresponding expression for $K^+ \ra \pi^+ \pi^+ \pi^- \gamma$ reads 
\beq \label{ano2}
-\frac{4 e G_8}{F^2} (N_{29} + N_{31})
\wt F_{\mu\nu} \partial^\mu K^+ \partial^\nu \pi^+ \pi^- \pi^-~.
\eeq
The decay $K_L \ra \pi^+ \pi^- \pi^0 \gamma$ receives a
contribution from 
\beq \label{ano3}
-\frac{2 e G_8}{F^2} (6 N_{28} + 3 N_{29} - 5 N_{30})
\wt F_{\mu\nu} \partial^\mu K_L  \partial^\nu \pi^0 
\pi^-  \pi^+~,
\eeq
and $K_S \ra \pi^+ \pi^- \pi^0 \gamma$ from 
\beq \label{ano4}
\frac{2 i e G_8}{F^2} \wt F_{\mu\nu} K_S 
[(5 N_{29} - N_{30} + 2 N_{31}) \partial^\mu \pi^0 
\pi^- \stackrel{\leftrightarrow}{\partial^\nu} \pi^+ -
2(N_{29} + N_{31}) \pi^0 \partial^\mu \pi^+ \partial^\nu \pi^-]~.
\eeq
Following the theoretical arguments given in \cite{BEP92}, the
coupling constants in the anomalous parity sector of $\cO(p^4)$
can be estimated as 
\beq
\ba{ll}
N_{28}^{\rm an} = \dfrac{a_1}{8\pi^2}~, \qquad \qquad &
N_{29}^{\rm an} = \dfrac{a_2}{32\pi^2}~, \\[10pt]
N_{30}^{\rm an} = \dfrac{3a_3}{16\pi^2}~, \qquad \qquad &
N_{31}^{\rm an} = \dfrac{a_4}{16\pi^2}~,
\ea  \label{Nan}
\eeq
where the dimensionless coefficients $a_i$ are expected to be
positive and of order one.

\begin{figure}
\centerline{\epsfig{file=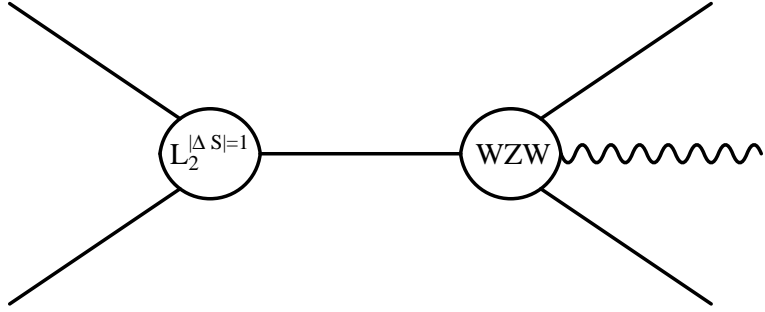,height=4cm}}
\caption{Reducible diagram contributing to the magnetic amplitude
at $\cO (p^4)$. A weak cubic vertex of $\cO (p^2)$ and an anomalous
vertex with three mesons and a photon are connected by a single
meson line.}
\label{fig:red}
\end{figure}

The second class of diagrams contributing to the magnetic amplitude are the
reducible ones (type iii). These amplitudes are due to diagrams with a single
meson line between a weak $|\Delta S| = 1$ vertex and an anomalous vertex
from the Wess--Zumino--Witten (WZW) functional \cite{WZW71}. For 
$K\to 3\pi\gamma$, all such diagrams have the structure shown in 
Fig.~\ref{fig:red}: a
weak cubic vertex and an anomalous vertex with three mesons and a photon.

In the case of $K^+ \ra \pi^0 \pi^0 \pi^+ \gamma$, there is only
one reducible contribution at $\cO(p^4)$: 
the kaon emits a neutral and a charged pion, where the $\pi^+$
subsequently makes an anomalous transition to $\pi^0 \pi^+
\gamma$,
$$
K^+ \stackrel{\scriptscriptstyle{\rm weak}}{\longrightarrow} \pi^0 (\pi^+
\stackrel{\scriptscriptstyle{\rm WZW}}{\longrightarrow} \pi^0 \pi^+ \gamma)~.
$$  
The corresponding amplitude is local because the $K^+\to \pi^0\pi^+$
vertex vanishes on--shell (remember that we are setting $G_{27}=0$ 
in direct emission amplitudes). 
Thus, the complete magnetic amplitude (adding the
direct term generated by (\ref{ano1})) takes the form
\beq
M^{\nu \rho \sigma}(K^+ \ra \pi^0 \pi^0 \pi^+ \gamma) = 
\frac{i G_8}{8 \pi^2 F^2} (3 a_2 - 6 a_3
-2) k^{\nu} p_3^{\rho} p_4^{\sigma}~.\label{MK00p}
\eeq

There are two types of reducible diagrams contributing to 
$K^+ \ra \pi^+ \pi^+ \pi^- \gamma$: the $K^+$ can make a weak 
transition into a real $\pi^+$ and a virtual $\pi^0$ (or
$\eta$) which is then transformed into a $\pi^+ \pi^-$ pair and a
photon,
$$
K^+ \stackrel{\scriptscriptstyle{\rm weak}}{\longrightarrow} \pi^+ (\pi^0
\stackrel{\scriptscriptstyle{\rm WZW}}{\longrightarrow} \pi^+ \pi^- \gamma)~,
$$
$$
K^+ \stackrel{\scriptscriptstyle{\rm weak}}{\longrightarrow} \pi^+ (\eta
\stackrel{\scriptscriptstyle{\rm WZW}}{\longrightarrow} \pi^+ \pi^- \gamma)~.
$$
The total magnetic amplitude is now given by 
\beqa
M^{\nu \rho \sigma}(K^+ \ra \pi^+ \pi^+ \pi^- \gamma) &=& 
\frac{i G_8}{2 \pi^2 F^2}  k^{\nu} p_3^{\rho} 
[(a_2 + 2 a_4) p_4^{\sigma} \no \\
&& + (M_{\eta}^2 - M_{K}^2) 
(\frac{p_1^{\sigma}}{s_{24}-M_{\eta}^2} +
\frac{p_2^{\sigma}}{s_{14}-M_{\eta}^2})]~,
\eeqa
with
\beqa
s_{14} &=& (p_1+p_4)^2 = \nu + t_3 +(M_K^2 + 3 M_{\pi}^2 - s)/2~, \no \\ 
s_{24} &=& (p_2+p_4)^2 = -\nu + t_3 +(M_K^2 + 3 M_{\pi}^2 - s)/2~. \no 
\eeqa 

For $K_L \ra \pi^+ \pi^- \pi^0 \gamma$ one may either contract
the anomalous $K_L K_S \pi^0 \gamma$ vertex with the weak $K_S \pi^+
\pi^-$ vertex, or the weak $K_L \ra \pi^+ \pi^-$ transition with the 
$\pi^+ \pi^- \pi^0 \gamma$ WZW vertex:
$$
K_L \stackrel{\scriptscriptstyle{\rm WZW}}{\longrightarrow} \pi^0 \gamma (K_S
\stackrel{\scriptscriptstyle{\rm weak}}{\longrightarrow} \pi^+ \pi^-)~,
$$
$$
K_L \stackrel{\scriptscriptstyle{\rm weak}}{\longrightarrow}
\pi^+ (\pi^-
\stackrel{\scriptscriptstyle{\rm WZW}}{\longrightarrow}
\pi^- \pi^0 \gamma)~,
$$
$$
K_L \stackrel{\scriptscriptstyle{\rm weak}}{\longrightarrow}
\pi^- (\pi^+
\stackrel{\scriptscriptstyle{\rm WZW}}{\longrightarrow}
\pi^+ \pi^0 \gamma)~.
$$ 
The last two diagrams give again a local amplitude for a similar 
reason\footnote{The on--shell amplitude for $K_L\to \pi^+\pi^-$ vanishes 
in the limit of CP conservation.}
as for $K^+ \ra \pi^0 \pi^0 \pi^+ \gamma$ in (\ref{MK00p}).
Together with the contribution from (\ref{ano3}),
we arrive at the magnetic amplitude
\beq
M^{\nu \rho \sigma}(K_L \ra \pi^+ \pi^- \pi^0 \gamma) = 
\frac{i G_8}{8 \pi^2 F^2} [24 a_1 + 3 a_2 - 30 a_3
-2 - \frac{4(M_K^2-M_\pi^2)}{s-M_K^2}] k^{\nu} p_3^{\rho} p_4^{\sigma}~.
\eeq

Finally, we turn to $K_S \ra \pi^+ \pi^- \pi^0 \gamma$. In this
case, the reducible diagrams have the following structure:
$$
K_S \stackrel{\scriptscriptstyle{\rm weak}}{\longrightarrow} \pi^+ (\pi^-
\stackrel{\scriptscriptstyle{\rm WZW}}{\longrightarrow} \pi^- \pi^0 \gamma)~,
$$
$$
K_S \stackrel{\scriptscriptstyle{\rm weak}}{\longrightarrow} \pi^- (\pi^+
\stackrel{\scriptscriptstyle{\rm WZW}}{\longrightarrow} \pi^+ \pi^0 \gamma)~,
$$
$$
K_S \stackrel{\scriptscriptstyle{\rm weak}}{\longrightarrow} \pi^0 (\pi^0
\stackrel{\scriptscriptstyle{\rm WZW}}{\longrightarrow} \pi^+ \pi^- \gamma)~,
$$
$$
K_S \stackrel{\scriptscriptstyle{\rm weak}}{\longrightarrow} \pi^0 (\eta
\stackrel{\scriptscriptstyle{\rm WZW}}{\longrightarrow} \pi^+ \pi^- \gamma)~.
$$
Combined with (\ref{ano4}), we obtain
\beqa
M^{\nu \rho \sigma}(K_S \ra \pi^+ \pi^- \pi^0 \gamma) &=& 
\frac{G_8}{8 \pi^2 F^2}  k^{\nu}  
\{(5 a_2 - 6 a_3 + 4 a_4 - 2) (p_2 - p_1)^\rho p_3^{\sigma}  \\
&& + 4 (M_K^2 - M_{\pi}^2) 
(\frac{p_2^{\rho}}{s_{14}-M_{\pi}^2} -
\frac{p_1^{\rho}}{s_{24}-M_{\pi}^2}) p_3^{\sigma} \no \\
&& + [ - 2a_2 - 4a_4+ \frac{4(M_K^2-M_\pi^2)}{s_{34}-M_\pi^2} + 
\frac{4(M_\eta^2-M_K^2)}{s_{34}-M_\eta^2}] p_1^{\rho} p_2^{\sigma}\}\no ~,
\eeqa
where
\beqa
s_{34} &=& (p_3+p_4)^2 = s + 2 (t_1 + t_2)~. \no 
\eeqa

\setcounter{equation}{0}
\setcounter{subsection}{0}
 
\section{Numerical results}
\label{sec:num}
Our numerical results for the  various channels 
are displayed in Tables \ref{K00P} -- \ref{KS}.
The first column shows the photon energy range. In the second column,
the contribution to the decay width generated by the generalized
bremsstrahlung amplitude $E_{\rm GB}$ in (\ref{EGB}) is listed, together 
with the corresponding errors due to the uncertainties of the $K \to 3 \pi$ 
parameters in (\ref{numval}). The next column shows the
relative change of the result if only the Low amplitude 
(\ref{Low}) is used instead of $E_{\rm GB}$. 
In the fourth column we see the effect of adding the electric
counterterms (using $k_f = 1$ in (\ref{counterterm1}) and
(\ref{KScount})) and the residual pion--loop contributions
$E_{\rm loop,subtracted}$ in (\ref{Efin}). 
$\Gamma_{\rm M}$ in the next column denotes the 
contribution to the decay width from the magnetic amplitudes 
(for $a_i = 1$); there is no interference between  electric and  magnetic 
amplitudes as long as the phase space integration is performed 
``symmetrically''.

For the branching ratios in the last column we distinguish
between the three channels where the leading--order amplitude is
not suppressed and the decay $K_S \ra \pi^+ \pi^- \pi^0 \gamma$ with
a suppressed bremsstrahlung amplitude. In the first group of transitions,
the  dominant $\cO(E_\gamma)$ effect is given by the difference
$\Gamma_{\rm GB}-\Gamma_{\rm Low}$. This deviation from 
Low's theorem, i.e. from a pure QED prediction, 
could possibly be observed in the near future. 
In the above channels, the residual pion--loop contribution suffers from 
relatively large theoretical uncertainties: the smallness of 
phase space amplifies isospin--breaking effects generated by the mass
difference $M_{\pi^0}-M_{\pi^+}$. However, the effect of 
$E_{\rm loop,subtracted}$ is always so small that it can hardly
be detected. The contribution of 
$E_{\rm counter}$,  evaluated within the factorization model,
is of the same order as  $E_{\rm loop,subtracted}$.
For $K^+\to\pi^0\pi^0\pi^+\gamma$ there is an almost 
complete destructive interference between loops and electric
counterterms, while for $K^+\to\pi^+\pi^+\pi^-\gamma$  we find
$E_{\rm counter}\simeq E_{\rm loop,subtracted}$. Finally,
in the $K_L$ channel $E_{\rm loop,subtracted}$ is bigger than
$E_{\rm counter}$ for large $E_\gamma$. For small $E_\gamma$,
the two amplitudes are comparable. Probably only large deviations from 
the naive expectation $k_f \sim \cO(1)$ could be observed.
Also the magnetic contribution is very much suppressed in these channels:
the ratio $\Gamma_{\rm M}/\Gamma_{\rm GB}$ is typically smaller than 
$10^{-3}$.

Interference effects between electric and magnetic amplitudes could in 
principle be larger. For instance, observables like 
${\rm det}(p_1, p_2, p_3, p_4)$ (for the decays with three different pions 
in the final state) or $\nu \, {\rm det}(p_1, p_2, p_3, p_4) $ (in the case
of two identical particles $\pi_1$, $\pi_2$) are sensitive to
such interferences. To $\cO(p^4)$, the interference term
\beq
\label{interference}
\ve^{\mu\nu\rho\sigma} (E_\mu  M_{\nu\rho\sigma}^*
+ E_\mu^*  M_{\nu\rho\sigma})
\eeq
is proportional to the relatively small absorptive part of the electric
amplitude. Thus, the leading--order piece of $E_\mu$ does not
contribute in (\ref{interference}). Nevertheless, the
possibility of interference measurements should be kept in mind
once sufficiently high statistics will have been achieved. 

For the three channels under consideration, the amplitude is
completely dominated by generalized bremsstrahlung. In the last column 
of Tables \ref{K00P} -- \ref{KL}, we therefore list the branching
ratios based on generalized bremsstrahlung only, corresponding to 
$\Gamma_{\rm GB}$ in the second column. The contributions to
the branching ratios from direct emission are completely concealed by 
the present experimental uncertainties of the $K\to 3 \pi$ parameters.

Within those errors, our predictions are consistent with standard
bremsstrahlung and with the available experimental results.
Our theoretical branching ratio for $K^- \ra \pi^0 \pi^0 \pi^- \gamma$
in Table \ref{K00P} for 
$E_{\gamma} > 10 {\rm MeV}$ can be compared directly with the experimental 
result \cite{Bolotov} 
\beq
\mbox{BR}(K^- \ra \pi^0 \pi^0 \pi^- \gamma) = (7.4^{+5.5}_{-2.9})
\cdot 10^{-6}~, \quad E_{\gamma} > 10 \; \mbox{MeV}~.
\eeq
For $K^+ \ra \pi^+ \pi^+ \pi^- \gamma$, Barmin et al. \cite{Barmin} have 
reported the branching ratio
\beq
\mbox{BR}(K^+ \ra \pi^+ \pi^+ \pi^- \gamma) = (1.10 \pm 0.48)
\cdot 10^{-4}~, \quad E_{\gamma} > 5 \; \mbox{MeV}~,
\eeq
to be compared with our theoretical prediction
\beq
{\rm BR}(K^+ \ra \pi^+ \pi^+ \pi^- \gamma) |_{\rm theor.} = (1.26 \pm 0.01)
\cdot 10^{-4}~, \quad E_{\gamma} > 5 \; \mbox{MeV}~,
\eeq
whereas Stamer et al. \cite{Stamer,PDG} have found
\beq
{\rm BR}(K^+ \ra \pi^+ \pi^+ \pi^- \gamma) = (1.0 \pm 0.4)
\cdot 10^{-4}~, \quad E_{\gamma} > 11 \; \mbox{MeV}~.
\eeq 

\begin{table}[t]
\caption{\protect{\label{K00P}} Numerical results 
for the decay $K^+ \ra \pi^0 \pi^0 \pi^+ \gamma$. 
The photon energy $E_{\gamma}$ and the decay widths
$\Gamma_{\rm GB}$, $\Gamma_{\rm M}$ are given in MeV.}  
$$
\begin{tabular}{|c|c|c|c|c|c|} \hline
$E_{\gamma}$  &
$\Gamma_{\rm GB}$ &
$\dfrac{\Gamma_{\rm GB}-\Gamma_{\rm Low}}{\Gamma}$ &
$\dfrac{\Gamma_{\rm E}-\Gamma_{\rm GB}}{\Gamma}$ &
$\Gamma_{\rm M}$ &
BR \\ \hline \hline
10--20 & $(1.38 \pm 0.02) \cdot 10^{-19}$ &
$1.4 \cdot 10^{-3}$  & $2.2 \cdot 10^{-5}$ & 
$2.8 \cdot 10^{-25}$ & $(2.60 \pm 0.03) \cdot 10^{-6}$ \\ \hline
20--30 & $(4.29 \pm 0.06) \cdot 10^{-20}$ &
$4.5 \cdot 10^{-3}$  & $8.2 \cdot 10^{-5}$  &
$7.5 \cdot 10^{-25}$  & $(8.05 \pm 0.01) \cdot 10^{-7}$ \\ \hline
30--40 & $(1.45 \pm 0.03) \cdot 10^{-20}$ &
$9.8 \cdot 10^{-3}$ & $2.5 \cdot 10^{-4}$ &
$1.1 \cdot 10^{-24}$  & $(2.72 \pm 0.05) \cdot 10^{-7}$ \\ \hline
40--50 & $(4.48 \pm 0.09) \cdot 10^{-21}$ &
$1.8 \cdot 10^{-2}$ & $2.2 \cdot 10^{-4}$ &
$1.1 \cdot 10^{-24}$  & $(8.42 \pm 0.18) \cdot 10^{-8}$ \\ \hline
50--60 & $(1.09 \pm 0.03) \cdot 10^{-21}$ &
$2.9 \cdot 10^{-2}$  & $-1.0 \cdot 10^{-3}$ &
$6.8 \cdot 10^{-25}$  & $(2.05 \pm 0.05) \cdot 10^{-8}$ \\ \hline
60--70 & $(1.49 \pm 0.05) \cdot 10^{-22}$ &
$4.3 \cdot 10^{-2}$  & $-6.8 \cdot 10^{-3}$ &
$2.0 \cdot 10^{-25}$  & $(2.81 \pm 0.09) \cdot 10^{-9}$ \\ \hline 
70--80 & $(3.48 \pm 0.12) \cdot 10^{-24}$ &
$5.6 \cdot 10^{-2}$  & $-1.9 \cdot 10^{-2}$ &
$8.9 \cdot 10^{-27}$  & $(6.55 \pm 0.23) \cdot 10^{-11}$ \\ \hline \hline
10--80 & $(2.01 \pm 0.03) \cdot 10^{-19}$ &
$3.3 \cdot 10^{-3}$  & $4.5 \cdot 10^{-5}$ &
$4.1 \cdot 10^{-24}$  & $(3.78 \pm 0.05) \cdot 10^{-6}$ \\ \hline
\end{tabular}
$$
\end{table}

\begin{table}
\caption{\protect{\label{KPPM}} Numerical results 
for the decay $K^+ \ra \pi^+ \pi^+ \pi^- \gamma$. }  
$$
\begin{tabular}{|c|c|c|c|c|c|} \hline
$E_{\gamma}$ &
$\Gamma_{\rm GB}$ &
$\dfrac{\Gamma_{\rm GB}-\Gamma_{\rm Low}}{\Gamma}$ &
$\dfrac{\Gamma_{\rm E}-\Gamma_{\rm GB}}{\Gamma}$ &
$\Gamma_{\rm M}$ &
BR \\ \hline \hline
10--20 & $(2.32 \pm 0.02) \cdot 10^{-18}$ &
$-1.7 \cdot 10^{-3}$ & $-4.2 \cdot 10^{-4}$ & 
 $1.3 \cdot 10^{-24}$  & $(4.36 \pm 0.04) \cdot 10^{-5}$ \\ \hline
20--30 & $(7.63 \pm 0.07) \cdot 10^{-19}$ &
$-4.8 \cdot 10^{-3}$  & $-1.2 \cdot 10^{-3}$ &
 $3.2 \cdot 10^{-24}$  & $(1.43 \pm 0.01) \cdot 10^{-5}$ \\ \hline
30--40 & $(2.62 \pm 0.03) \cdot 10^{-19}$ &
$-9.2 \cdot 10^{-3}$  & $-2.4 \cdot 10^{-3}$ &
 $4.1 \cdot 10^{-24}$  & $(4.93 \pm 0.05) \cdot 10^{-6}$ \\ \hline
40--50 & $(7.66 \pm 0.08) \cdot 10^{-20}$ &
$-1.5 \cdot 10^{-2}$  & $-4.1 \cdot 10^{-3}$ &
 $3.2 \cdot 10^{-24}$  & $(1.44 \pm 0.01) \cdot 10^{-6}$ \\ \hline
50--60 & $(1.43 \pm 0.02) \cdot 10^{-20}$ &
$-2.1 \cdot 10^{-2}$  & $-6.2 \cdot 10^{-3}$ &
 $1.3 \cdot 10^{-24}$  & $(2.69 \pm 0.03) \cdot 10^{-7}$ \\ \hline
60--70 & $(7.23 \pm 0.09) \cdot 10^{-22}$ &
$-2.8 \cdot 10^{-2}$  & $-8.5 \cdot 10^{-3}$ &
 $1.2 \cdot 10^{-25}$  & $(1.36 \pm 0.02) \cdot 10^{-8}$ \\ \hline \hline
10--70 & $(3.44 \pm 0.03) \cdot 10^{-18}$ &
$-3.4 \cdot 10^{-3}$  & $-8.5 \cdot 10^{-4}$ &
 $1.3 \cdot 10^{-23}$  & $(6.46 \pm 0.06) \cdot 10^{-5}$ \\ \hline
\end{tabular}
$$
\end{table}

\begin{table}[t]
\caption{\protect{\label{KL}} Numerical results 
for the decay $K_L \ra \pi^+ \pi^- \pi^0 \gamma$.}  
$$
\begin{tabular}{|c|c|c|c|c|c|} \hline
$E_{\gamma}$ &
$\Gamma_{\rm GB}$ &
$\dfrac{\Gamma_{\rm GB}-\Gamma_{\rm Low}}{\Gamma}$ &
$\dfrac{\Gamma_{\rm E}-\Gamma_{\rm GB}}{\Gamma}$ &
$\Gamma_{\rm M}$ &
BR \\ \hline \hline
10--20 & $(1.32 \pm 0.02) \cdot 10^{-18}$ &
$-4.5 \cdot 10^{-3}$  & $-4.2 \cdot 10^{-4}$ & 
$2.1 \cdot 10^{-26}$ & $(1.04 \pm 0.02) \cdot 10^{-4}$ \\ \hline
20--30 & $(4.89 \pm 0.07) \cdot 10^{-19}$ &
$-1.3 \cdot 10^{-2}$  & $-1.1 \cdot 10^{-3}$ &
$4.4 \cdot 10^{-26}$  & $(3.84 \pm 0.06) \cdot 10^{-5}$ \\ \hline
30--40 & $(1.98 \pm 0.03) \cdot 10^{-19}$ &
$-2.5 \cdot 10^{-2}$  & $-1.8 \cdot 10^{-3}$ &
$5.0 \cdot 10^{-26}$  & $(1.55 \pm 0.03) \cdot 10^{-5}$ \\ \hline
40--50 & $(7.33 \pm 0.11) \cdot 10^{-20}$ &
$-4.0 \cdot 10^{-2}$  & $-1.7 \cdot 10^{-3}$ &
$3.7 \cdot 10^{-26}$  & $(5.76 \pm 0.10) \cdot 10^{-6}$ \\ \hline
50--60 & $(2.13 \pm 0.04) \cdot 10^{-20}$ &
$-5.8 \cdot 10^{-2}$  & $2.5 \cdot 10^{-3}$ &
$1.7 \cdot 10^{-26}$  & $(1.67 \pm 0.03) \cdot 10^{-6}$ \\ \hline
60--70 & $(3.39 \pm 0.06) \cdot 10^{-21}$ &
$-7.9 \cdot 10^{-2}$  & $2.7 \cdot 10^{-2}$ &
$3.5 \cdot 10^{-27}$  & $(2.67 \pm 0.05) \cdot 10^{-7}$ \\ \hline
70--80 & $(8.04 \pm 0.15) \cdot 10^{-23}$ &
$-9.9 \cdot 10^{-2}$  & $2.2 \cdot 10^{-1}$ &
$9.5 \cdot 10^{-29}$  & $(6.32 \pm 0.13) \cdot 10^{-9}$ \\ \hline \hline
10--80 & $(2.11 \pm 0.03) \cdot 10^{-18}$ &
$-1.0 \cdot 10^{-2}$  & $-6.6 \cdot 10^{-4}$ &
$1.7 \cdot 10^{-25}$  & $(1.65 \pm 0.03) \cdot 10^{-4}$ \\ \hline
\end{tabular}
$$
\end{table}
\begin{table}
\caption{\protect{\label{KS}} Numerical results 
for the decay $K_S \ra \pi^+ \pi^- \pi^0 \gamma$.}  
$$
\begin{tabular}{|c|c|c|c|c|c|} \hline
$E_{\gamma}$ &
$\Gamma_{\rm GB}$ &
$\dfrac{\Gamma_{\rm GB}-\Gamma_{\rm Low}}{\Gamma}$ &
$\Gamma_{\rm E}$ &
$\Gamma_{\rm M}$ &
BR \\ \hline \hline
10--20 & $(1.29 \pm 0.34) \cdot 10^{-21}$ &
$1.2 \cdot 10^{-2}$  & $1.1 \cdot 10^{-21}$ & 
 $6.5 \cdot 10^{-25}$  & $1.5 \cdot 10^{-10}$  \\ \hline
20--30 & $(5.15 \pm 1.28) \cdot 10^{-22}$ &
$3.8 \cdot 10^{-2}$  &  $3.4 \cdot 10^{-22}$  &
 $1.6 \cdot 10^{-24}$  & $4.7 \cdot 10^{-11}$  \\ \hline
30--40 & $(2.34 \pm 0.53) \cdot 10^{-22}$ &
$7.7 \cdot 10^{-2}$  & $9.7 \cdot 10^{-23}$  &
 $2.3 \cdot 10^{-24}$  & $1.3 \cdot 10^{-11}$  \\ \hline
40--50 & $(9.97 \pm 2.12) \cdot 10^{-23}$ &
$1.2 \cdot 10^{-1}$  & $2.0 \cdot 10^{-23}$  &
 $2.1 \cdot 10^{-24}$  & $2.9 \cdot 10^{-12}$  \\ \hline
50--60 & $(3.34 \pm 0.68) \cdot 10^{-23}$ &
$1.6 \cdot 10^{-1}$  & $2.2 \cdot 10^{-24}$  &
 $1.2 \cdot 10^{-24}$  & $4.6 \cdot 10^{-13}$  \\ \hline
60--70 & $(6.09 \pm 1.22) \cdot 10^{-24}$ &
$2.1 \cdot 10^{-1}$  & $2.3 \cdot 10^{-25}$  &
 $3.4 \cdot 10^{-25}$  & $7.8 \cdot 10^{-14}$  \\ \hline
70--80 & $(1.62 \pm 0.32) \cdot 10^{-25}$ &
$2.4 \cdot 10^{-1}$  & $1.7 \cdot 10^{-26}$  &
 $1.2 \cdot 10^{-26}$  & $4.0 \cdot 10^{-15}$  \\ \hline \hline
10--80 & $(2.18 \pm 0.55) \cdot 10^{-21}$ &
$3.3 \cdot 10^{-2}$  & $1.6 \cdot 10^{-21}$  &
 $8.2 \cdot 10^{-24}$  & $2.2 \cdot 10^{-10}$  \\ \hline
\end{tabular}
$$
\end{table}

For the decay $K_S \ra \pi^+ \pi^- \pi^0 \gamma$ the situation
is quite different. To lowest chiral order, the amplitude can only
proceed through a $\Delta I=3/2$ transition (via bremsstrahlung)
and is therefore suppressed by the $\Delta I = 1/2$ rule.
Consequently, the next--to--leading order contributions generated by
octet operators are becoming relatively more important\footnote{A 
similar phenomenon occurs in 
the  $K^+\to \pi^+\pi^0\gamma$ decay \cite{ENP94,DI95}.}. 
At the one--loop level, two--pion intermediate states do not contribute.
Therefore, the $\cO(p^4)$ part of the electric amplitude is
essentially determined by the counterterm
\beq
N_{K_S}(\mu) := \left[7(N^r_{14}-N^r_{16})+5( N^r_{15}+ N_{17})\right](\mu)
\eeq
that is predicted to be large by the factorization model in 
(\ref{KScount}). Its rather modest scale dependence,
\beq
N_{K_S}(\mu_2) = N_{K_S}(\mu_1) + \frac{3}{8\pi^2}\mbox{ln}(\mu_1 / \mu_2)~,
\eeq
is compensated by loop graphs with kaon intermediate states which
we have neglected. The uncertainty induced by this scale dependence,
\beq
\frac{ N_{K_S}(0.5\mbox{ GeV}) - N_{K_S}(1\mbox{ GeV}) }{
N_{K_S}^{\rm FM} } \simeq 0.09~,
\eeq
with $N_{K_S}^{\rm FM}$ given by (\ref{KScount}) for $k_f=1$,
is certainly smaller than the intrinsic uncertainty of the factorization
hypothesis. 

The corresponding numerical results 
are displayed in Table \ref{KS}. The numbers for
$\Gamma_{\rm E}$ are obtained from the sum of $E_{\rm GB}$ (using
the central values of the input parameters $b_2$ and $d_2$ in 
(\ref{numval})) and the aforementioned $\cO(p^4)$ counterterm amplitude. 
Note that the interference is destructive and especially pronounced
at large values of $E_\gamma$.
The contribution of the magnetic amplitude is again shown for 
$a_i = 1$. For this channel we list the total branching ratio $\mbox{BR}
= (\Gamma_{\rm E} + \Gamma_{\rm M}) / \Gamma_{\rm tot}(K_S)$ for
the various photon energy bins. We
do not give errors for these branching ratios because, unlike
for the other three channels, the direct emission amplitude matters
with unknown theoretical uncertainties (factorization model).

Remembering the projected DA$\Phi$NE yield of $7.5\times 10^9 K_L K_S$ 
pairs per year, the $K_S \ra \pi^+ \pi^-
\pi^0 \gamma$ decay rate is still too small for the coming
generation of kaon experiments. With an additional improvement of
statistics, some information might be achieved via time--interference 
measurements \cite{DP} ($K_{L,S}\to \pi^+\pi^-\pi^0\gamma$) similar to those  
recently performed in the non--radiative case \cite{E621,CPLEAR}.
Then interference effects between electric and magnetic amplitudes could in 
principle be measured since a term like
\beq
\label{interference2}
\ve^{\mu\nu\rho\sigma} (E_\mu  M_{\nu\rho\sigma}^*
- E_\mu^*  M_{\nu\rho\sigma})
\eeq
is generated. In contrast to (\ref{interference}), this term
is proportional to the leading--order piece of $E_\mu$.
We stress that even fixed--target experiments,
through regeneration, can perform time--interference measurements
and in this case a larger statistics is expected.
Thus, the $K_S \ra \pi^+ \pi^-\pi^0 \gamma$
decay mode may still turn out to be a valuable
probe for kaon physics parameters that is not drowned by bremsstrahlung.

%

\setcounter{equation}{0}
\setcounter{subsection}{0}
\section{Conclusions}
\label{sec:conc}
Anticipating substantial improvements in the statistics of 
$K\to 3 \pi\gamma$ decays in the near future, we have performed a
comprehensive and complete analysis of these decays to $\cO(p^4)$ in
the low--energy expansion of the Standard Model. To lowest order,
$\cO(p^2)$, the decay amplitudes are determined by the corresponding
non--radiative amplitudes via Low's theorem (bremsstrahlung). At 
next--to--leading order, there are different contributions to both
electric and magnetic parts of the amplitudes: loops and tree--level
(counterterm) amplitudes, reducible and irreducible contributions.

A major aspect of our analysis is the concept of ``generalized bremsstrahlung"
that transfers the available theoretical or experimental information on
$K\to 3 \pi$ decays to the corresponding radiative amplitudes in an
optimal way at the level of $\cO(p^4)$. For the numerical analysis,
we have used the factorization hypothesis to estimate the relevant 
low--energy constants. 

Returning to the three issues addressed in the introduction, we may
summarize our findings as follows:
\ben
\item[i.]
In all three channels where the leading--order amplitudes are not
suppressed ($K^+\to\pi^0\pi^0\pi^+\gamma$, $K^+\to\pi^+\pi^+\pi^-\gamma$, 
$K_L\to\pi^+\pi^-\pi^0\gamma$), generalized bremsstrahlung
completely dominates the amplitudes to $\cO(p^4)$. The differences to
the QED prediction (standard or internal bremsstrahlung) could be 
experimentally observed in the forthcoming round of kaon experiments, 
at least from the statistical point of view.
\item[ii.] 
For the same channels, it will hardly be possible to extract the
appropriate combinations of low--energy constants from experiment in the
near future. This conclusion hinges, of course, on the assumption that
the factorization estimates are not off by an order--of--magnitude
in amplitude. In contrast, the counterterm amplitude is important for
$K_S\to\pi^+\pi^-\pi^0\gamma$, especially if the rather large factorization
estimate is reliable. However, for this decay mode the branching ratio
is probably too small to be detected soon.
\item[iii.]
As a general conclusion, the Standard Model allows for quite definite 
predictions for radiative kaon decays into three pions. Especially for
$K^+\to\pi^0\pi^0\pi^+\gamma$, $K^+\to\pi^+\pi^+\pi^-\gamma$ and
$K_L\to\pi^+\pi^-\pi^0\gamma$, the accuracy of these predictions is
at the moment only limited by the precision with which the parameters
of the non--radiative decay amplitudes are known. For 
$K_S\to\pi^+\pi^-\pi^0\gamma$, there is some theoretical uncertainty
related to the relevant low--energy constants.
\een

As soon as more accurate data will lead to better precision for the
$K\to 3 \pi$ parameters, the predictions of the radiative amplitudes
can be improved accordingly. Although we have only considered total
rates and photon energy spectra in this analysis, the investigation
of more subtle effects like the interference between electric and
magnetic amplitudes may then become feasible.

%
\section*{Acknowledgements}
\noindent
One of us (G. D'A.) wants to thank F. Sannino for discussions. 

\appendix{\section*{Appendix: Loop amplitudes}}
\newcounter{zahler}
\renewcommand{\thesection}{\Alph{zahler}}
\renewcommand{\theequation}{\Alph{zahler}.\arabic{equation}}
\setcounter{zahler}{1}
\setcounter{equation}{0}
\label{app:loop}
In this Appendix, we collect the main results of Ref.~\cite{DEIN96a}
for the calculation of loop amplitudes corresponding to the diagram
in Fig.~\ref{fig:loop}.

First, we calculate the loop amplitude for the non--radiative process
$K\to 3\pi$. In our case, $-p_a=-p_4$ is the kaon momentum and $V_1$,
$V_2$ are nonleptonic weak and strong vertices, respectively. 
The pion momenta are generically denoted $p_b,p_c,p_d$.

We characterize the vertices $V_1,V_2$ in momentum space 
by constants $a_i,b_i$:
\beqa
\label{vertex}
V_1 &=& a_0 + a_1 p_a \czdot p_b + a_2 p_a \czdot x + a_3(x^2 - M_x^2)
+ a_4(y^2 - M_y^2) + a_5(p_a^2 - M_a^2) + a_6(p_b^2 - M_b^2) \no \\
V_2 &=& b_0 + b_1 p_c \czdot p_d + b_2 p_c \czdot x + b_3(x^2 - M_x^2)
+ b_4(y^2 - M_y^2) + b_5(p_c^2 - M_c^2) + b_6(p_d^2 - M_d^2)~.\no \\
\eeqa
With $P = p_c + p_d$, the non--radiative loop amplitude of Fig.~\ref{fig:loop}
can be represented in the following form (all external lines
are on--shell):
\beqa
\label{Floop}
F(P) &=& A(M_x) [a_1 b_4 p_a \czdot p_b + a_4 b_1 p_c \czdot p_d +
a_4 b_4 (P^2 + M_x^2 - M_y^2) + a_0 b_4 + a_4 b_0] \no \\
&& \mbox{} + A(M_y) [a_1 b_3 p_a \czdot p_b + a_2 b_3 p_a \czdot P +
a_3 b_1 p_c \czdot p_d + a_3 b_2 p_c \czdot P \no \\
&& \hspace*{1.8cm} + a_3 b_3(P^2 - M_x^2 + M_y^2) + a_0 b_3 + a_3 b_0] 
\no \\
&& \mbox{} + B(P^2,M_x,M_y)[a_0 b_0 + a_0 b_1 p_c \czdot p_d +
a_1 b_0 p_a \czdot p_b + a_1 b_1 p_a \czdot p_b p_c \czdot p_d] \no \\
&& \mbox{} + B_1(P^2,M_x,M_y)[a_0 b_2 p_c \czdot P + a_2 b_0 p_a \czdot P
+ a_1 b_2 p_a \czdot p_b p_c \czdot P + a_2 b_1 p_c \czdot p_d p_a \czdot P] 
\no \\
&& \mbox{} + a_2 b_2[p_a \czdot p_c B_{20}(P^2,M_x,M_y) + p_a \czdot P
p_c \czdot P B_{22}(P^2,M_x,M_y)]~.
\eeqa
The various functions in (\ref{Floop}) are as defined conventionally
(in $d$ dimensions):
\beqa
\label{ABdef}
A(M) &=& \frac{1}{i} \int \frac{d^d x}{(2\pi)^d} 
\frac{1}{x^2 - M^2} \no \\
(B,B_1 P_\mu,g_{\mu\nu} B_{20} + P_\mu P_\nu B_{22}) &=&
\frac{1}{i} \int \frac{d^d x}{(2\pi)^d}
\frac{(1,x_\mu,x_\mu x_\nu)}{(x^2 - M_x^2)[(x-P)^2 - M_y^2]}~.
\eeqa

We have chosen to express $F(P)$ in terms of the scalar products
\beq
\label{scalar}
p_a \czdot p_b, \quad p_c \czdot p_d, \quad P^2, \quad
p_a \czdot P, \quad p_c \czdot P, \quad p_a \czdot p_c
\eeq
instead of using kinematical relations to express all scalar products
in terms of the two independent scalar variables $s,\nu$.
Note that the analytically non--trivial part of (\ref{Floop}), involving
the various $B$ functions, contains only the on--shell couplings
$a_0,a_1,a_2,b_0,b_1,b_2$. The off--shell couplings $a_3,a_4,b_3,b_4$
appear only together with the divergent constants $A(M)$. Since
these terms are polynomials in the momenta of at most degree two,
they will enter in the radiative amplitude only through internal
bremsstrahlung and will therefore eventually be absorbed in
$E^\mu_{\rm GB}$. The on--shell coefficients for the various
channels are listed in Table \ref{tab:coeff}.

\begin{table}[t]
\label{tab:coeff}
\caption{Coefficients of the vertices $V_1,V_2$ defined in 
(\protect\ref{vertex}) for the various loop diagrams. Only the
relevant on--shell coefficients are listed.}
\[ \ba{|c||c|c|c||c|c|c|} \hline
 K(-p_a) \to \pi (p_b) ~+   &&&&&&\\ \pi(x)\pi(y)\to \pi(p_c)\pi(p_d) 
& a_0 & a_1 & a_2 & b_0 & b_1 & b_2  \\ \hline
 K^+ \to \pi^+ ~+ &&&&&&\\ \pi^+\pi^- \to \pi^+\pi^-  
& - 2 M^2_K & -2  & -2  &  2 M^2_\pi & 2  & -2 \\ \hline
 K^+ \to \pi^+ ~+ &&&&&&\\ \pi^+\pi^- \to \pi^0\pi^0 
& - 2 M^2_K & -2  & -2  &  M^2_\pi & 2  & 0  \\ \hline
 K^+ \to \pi^+ ~+ &&&&&&\\ \pi^0\pi^0 \to \pi^+\pi^- 
& -  M^2_K & -2  & 0  &  M^2_\pi & 2  & 0 \\ \hline
 K^+ \to \pi^- ~+ &&&&&&\\ \pi^+\pi^+ \to \pi^+\pi^+
&        0 &  2  & 0  &  0       & -2  & 0 \\ \hline
 K^+ \to \pi^0 ~+ &&&&&&\\ \pi^+\pi^0 \to \pi^+\pi^0
& -  M^2_K & 0   & -2  &  M^2_\pi & 0  & -2 \\ \hline
 K^0 \to \pi^+ ~+ &&&&&&\\ \pi^0\pi^- \to \pi^0\pi^-
&    M^2_K/\sqrt{2} & 0  & \sqrt{2}  &  M^2_\pi & 0 
                              & -2 \\ \hline
 K^0 \to \pi^- ~+ &&&&&&\\ \pi^0\pi^+ \to \pi^0\pi^+
&    M^2_K/\sqrt{2} & 0  & \sqrt{2}  &  M^2_\pi & 0  
                              & -2 \\ \hline
 K^0 \to \pi^0 ~+ &&&&&&\\ \pi^+\pi^- \to \pi^+\pi^-  
&    M^2_K/\sqrt{2} & \sqrt{2}  & 0  & 2 M^2_\pi & 2  
                              & -2 \\ \hline
 \ea    \]
\end{table}                         

We now turn to the radiative loop amplitude and decompose it into
two parts:
\beq
E^\mu_{\rm loop} = G^\mu + H^\mu . \label{decomp}
\eeq
The amplitude $G^\mu$ can be expressed through derivatives of the
non--radiative loop amplitude $F$ in (\ref{Floop}) with respect to the
various scalar products (\ref{scalar}). In some of the following terms, the
momentum $P$ has to be replaced by $P + k$, leaving all scalar products
unchanged that do not contain $P$ explicitly:
\beqa
\label{Gmu}
G^\mu &=& F(P) \Sigma^\mu + \frac{F(P+k) - F(P)}{k \czdot P} \Lambda^\mu_{cd}
+ \frac{\partial F}{\partial(p_a \czdot p_b)}(P) \Lambda^\mu_{ab} \no \\
&& \mbox{} + \frac{\partial F}{\partial(p_a \czdot P)}(P)\Lambda^\mu_{aP} +
\frac{\partial F}{\partial(p_c \czdot p_d)}(P+k)\Lambda^\mu_{cd} 
+ \frac{\partial F}{\partial (p_c \czdot P)}(P+k)\Lambda^\mu_{cP} \no \\
&& \mbox{} + \left[ q_a t_c \frac{\partial F}{\partial(p_a \czdot p_c)}(P)
- q_c t_a \frac{\partial F}{\partial(p_a \czdot p_c)}(P+k)\right]
D^\mu_{ac} \no \\
&& \mbox{} - \frac{1}{2}(q_c + q_d) t_a t_c \left[ 
\frac{\partial^2 F}{\partial(p_a \czdot P)\partial(p_c \czdot P)}(P)
D^\mu_{aP} - \frac{\partial^2 F}{\partial(p_a \czdot P)\partial(p_c \czdot
P)}(P+k) D^\mu_{cP} \right].\no \\
\eeqa
We have used the definitions (\ref{defs}). When $P$ appears as an index
(e.g., in $\Lambda^\mu_{aP}$ or $D^\mu_{cP}$), the corresponding momentum 
and charge in (\ref{defs}) are $P$ and $q_c + q_d$, respectively. 

The second part $H^\mu$ of the loop amplitude (\ref{decomp}) cannot
be expressed in terms of $F$ or derivatives thereof. 
For the relevant case of equal loop masses ($M_x = M_y = M_\pi =: M$),
$H^\mu$ takes on the following compact form:
\beqa
\label{Hmu}
H^\mu &=& a_2(t_b p^\mu_a - t_a p^\mu_b) \{(q_x-q_y)(2b_0 + 2b_1 p_c \czdot
p_d +b_2 p_c \czdot P) \wt{C_{20}}(P^2, - k \czdot P) \no \\*
&& \mbox{} + b_2(q_x+q_y) [-2p_c \czdot P
\wt{C_{31}}(P^2,-k \czdot P) + 2t_c \wt{C_{32}}(P^2,-k \czdot P) 
- p_c \czdot P \wt{C_{20}}(P^2,-k \czdot P)]\} \no \\
&& \mbox{} + b_2(t_d p^\mu_c - t_c p^\mu_d) \{(q_x - q_y)
[2a_0 + 2a_1 p_a \czdot p_b + a_2(p_a \czdot P + t_a)] 
\wt{C_{20}}((P+k)^2, k \czdot P) \no \\
&& \mbox{} + a_2(q_x + q_y) [-2(p_a \czdot P + t_a)
\wt{C_{31}}((P+k)^2,k \czdot P)  - 2t_a \wt{C_{32}}((P+k)^2,
k \czdot P)\no \\ 
&& \mbox{} - (p_a \czdot P + t_a) \wt{C_{20}}((P+k)^2,k \czdot P)]\}~.
\eeqa
The functions $\wt{C_{ij}}$ are defined as
\beq
\wt{C_{ij}}(u,v) = \frac{C_{ij}(u,v) - C_{ij}(u,0)}{v} \label{Cdiff}
\eeq
in terms of the three--propagator one--loop functions 
$C_{ij}(p^2,k \czdot p)$ for $k^2 = 0$:
\beqa
\label{Cdef}
\lefteqn{ \frac{1}{i} \int \frac{d^dx}{(2\pi)^d} \frac{ \{x_\mu x_\nu,
x_\mu x_\nu x_\rho\}}{(x^2 - M^2)[(x+p)^2 - M^2][(x+k)^2 - M^2]} = } \no \\
&=& \{ C_{20}(p^2,k\czdot p)g_{\mu\nu} + \ldots, C_{31}(p^2,k \czdot p)
(p_\mu g_{\nu\rho} + p_\nu g_{\mu\rho} + p_\rho g_{\mu\nu}) \no \\
&& \mbox{} + C_{32}(p^2,k \czdot p)(k_\mu g_{\nu\rho} + k_\nu g_{\mu\rho}
+ k_\rho g_{\mu\nu}) + \ldots\}.
\eeqa

We recall the following observations from Ref.~\cite{DEIN96a}:
\begin{enumerate}
\item[i.] The amplitudes $G^\mu$ in (\ref{Gmu}) and $H^\mu$ in
(\ref{Hmu}) are separately gauge invariant.
\item[ii.] The amplitude $H^\mu$ is finite and at least of $\cO(k)$.
It only contains the on--shell couplings
$a_0,a_1,a_2,b_0,b_1,b_2$ defined in (\ref{vertex}) and the charges
$q_x,q_y$ of the particles in the loop. 
\item[iii.] The generalized bremsstrahlung part of the loop amplitude
is contained in $G^\mu$. Denoting by  $E^\mu_{\rm GB}({\rm loop})$  
the result obtained by inserting for $A(s,\nu)$ the on--shell loop 
amplitude (\ref{Floop}) in Eq.~(\ref{EGB}), the difference
\beq
\Delta^\mu = G^\mu - E^\mu_{\rm GB}({\rm loop})  
\eeq
is at least of $\cO(k)$. Moreover, by construction of $E^\mu_{\rm GB}$
all the divergences in $\Delta^\mu$ are renormalized by counterterms 
with an explicit field strength tensor.
Finally, $\Delta^\mu$ is finite for $a_2 b_2 = 0$.
\end{enumerate}

Putting everything together, the subtracted loop amplitude 
$E^\mu_{\rm loop,subtracted}$ in (\ref{Efin}) is given by
\beq
E^\mu_{\rm loop,subtracted} =  \sum_{\rm loops} (\Delta^\mu +H^\mu)~.
\label{Esub}
\eeq
The sum extends over the various configurations listed in Table
\ref{tab:coeff}.


\end{document}